\documentclass[aps,prb,groupedaddress,amsmath,amssymb,twocolumn]{revtex4}

\usepackage{amssymb}
\usepackage{amsmath}
\usepackage{graphicx}
\usepackage{subfigure}
\usepackage{textcomp}
\usepackage{color}
\usepackage{amsfonts}
\usepackage{epsfig}
\usepackage{hyperref}

\begin{document}
\title{Tunable point nodes from line node semimetals via application of light}

\author{Awadhesh Narayan}
\email{awadhesh@illinois.edu}
\affiliation{Department of Physics, University of Illinois at Urbana-Champaign, Urbana, Illinois, USA.}

\date{\today}

\begin{abstract}
We propose that illumination with light provides a useful platform for creating tunable semimetals. We show that by shining light on semimetals with a line degeneracy, one can convert them to a point node semimetal. These point nodes are adjustable and their position can be controlled by simply rotating the incident light beam. We also discuss the implications of this change in Fermi surface topology, as manifested in transport observables.
\end{abstract}

\maketitle

Line node semimetals are a class of topological semimetals, which have caught wide attention recently. In these gapless systems, bands touch along lines resulting in quite peculiar loop-like Fermi surfaces~\cite{volovik2003universe,burkov2011topological}. Such symmetry protected topological nodal line semimetals can exist both in presence as well as absence of spin orbit coupling~\cite{fang2015topological}. Moreover, they harbor nearly flat surface states, which could have intriguing properties~\cite{heikkila2015flat}.

In a remarkably rapid flurry of experimental and theoretical activity, a number of materials have been proposed to host line nodes. These include carbon systems~\cite{weng2015topological,PhysRevLett.115.026403,chen2015nanostructured}, copper nitrides~\cite{PhysRevLett.115.036806,PhysRevLett.115.036807}, non centrosymmetric tantalum compounds~\cite{ali2014noncentrosymmetric,bian2016topological}, phosphides~\cite{xie2015new}, zirconium based compounds~\cite{schoop2015dirac,neupane2016observation}, a platinum-tin compound~\cite{wu2016dirac}, as well as elemental solids such as body centred cubic iron~\cite{gosalbez2015chiral} and face centred cubic calcium~\cite{hirayama2016topological}.

At the same time there has been a growing interest in using light to manipulate the electronic states of a diverse set of materials~\cite{wang2013observation,sie2015valley,kim2014ultrafast}. On the theoretical front this has led to the proposal of floquet topological insulators~\cite{oka2009photovoltaic,lindner2011floquet}. Application of circularly polarized light to low-dimensional materials, including graphene~\cite{kitagawa2011transport,gu2011floquet,perez2014floquet} and silicene~\cite{ezawa2013photoinduced}, has been suggested as a means to generate topologically non-trivial gaps in their spectra. Light induced floquet dynamics has also been investigated for Dirac-Weyl semimetals~\cite{wang2014floquet,narayan2015floquet}. Intriguing consequences of applying light to gapless systems are now beginning to be explored: photoinduced anomalous Hall effect in Weyl semimetals~\cite{chan2016chiral}, chiral pumping in Dirac semimetals~\cite{PhysRevB.93.155107}, macroscopic chiral currents on surfaces of Dirac and Weyl semimetals~\cite{PhysRevLett.116.156803}, to name a few. There have also been proposals to floquet engineer gapless semimetallic phases in various systems~\cite{bomantara2015floquet,PhysRevB.93.144114,zou2016floquet}.

Motivated by these developments, in this Rapid Communication we propose that by applying light of suitable polarization and strength one can drive a Lifshitz transition in line node semimetals. This allows one to create tunable point nodes from line node semimetals, where the position of the point nodes can be engineered by simply rotating the laser beam. We also discuss the consequences of such a change in Fermi surface topology and band structure as manifested in various observables.

\begin{figure*}[t]
\begin{center}
  \includegraphics[scale=0.55]{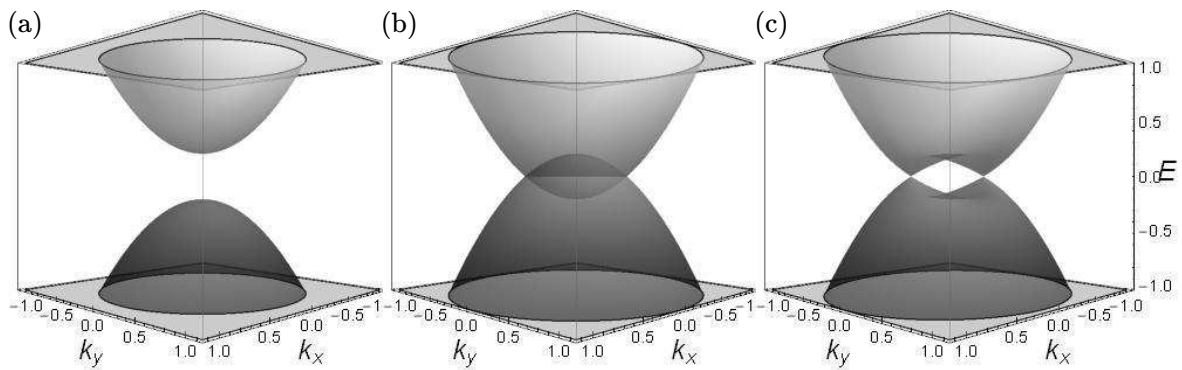}
  \caption{Energy eigenvalues for the low-energy model with $k_{z}=0$ for (a) $\Delta\epsilon=0.2$, (b) $\Delta\epsilon=-0.2$, in the absence of light. Note the line node degeneracy in the latter case and a band insulator in the former. (c) Photon dressed energy eigenvalues with $\Delta\epsilon=-0.2$, $\mathcal{A}_{y}=\mathcal{A}_{z}=0.5$ for a circularly polarized beam. The line degeneracy between the two bands is reduced to two points located symmetrically along $k_{y}=0$. Here and henceforth, we have set $\epsilon_{0}=0$, $a_{xy}=0$, $a_{z}=1.0$, $v=1.0$, $b_{xy}=1.0$, $b_{z}=1.0$ and $\omega=1.0$.  }  \label{bands}
\end{center}
\end{figure*}

Let us begin by considering the low-energy Hamiltonian for line node semimetals proposed by Kim and coauthors~\cite{PhysRevLett.115.036806} 

\begin{eqnarray}
 H&=&[\epsilon_{0}+a_{xy}(k_{x}^{2}+k_{y}^{2})+a_{z}k_{z}^{2}]I + vk_{z}\sigma_{y} \nonumber \\
  &+& [\Delta\epsilon+b_{xy}(k_{x}^{2}+k_{y}^{2})+b_{z}k_{z}^{2}]\sigma_{z},
\end{eqnarray}

\noindent where $\sigma_{i}$ ($i=x,y,z$) is the triad of Pauli matrices and $I$ is the identity matrix. The energy eigenvalues are given by $E=\epsilon_{0}+a_{xy}(k_{x}^{2}+k_{y}^{2})+a_{z}k_{z}^{2} \pm \sqrt{[\Delta\epsilon+b_{xy}(k_{x}^{2}+k_{y}^{2})+b_{z}k_{z}^{2}]^{2}+v^{2}k_{z}^{2}}$. If $\Delta\epsilon >0$, then the model yields an insulating band structure (for $b_{xy}>0$). On the other hand for $\Delta\epsilon <0$, the two bands form a line node degeneracy at $k_{z}=0$ and $k_{x}^{2} + k_{y}^{2} = -\Delta\epsilon/b_{xy}$. For this model, we have inversion, $\mathcal{P}=\sigma_{z}$ and time reversal, $\mathcal{T}=\mathcal{K}$, where $\mathcal{K}$ denotes complex conjugation. The band structures corresponding to the above two cases are shown for $k_{z}=0$ in Fig.~\ref{bands}(a) and (b), respectively. Notice that for $\Delta\epsilon <0$ [Fig.~\ref{bands}(b)] the two bands are degenerate along a circle forming a line node semimetal. 

\begin{figure}[b]
\begin{center}
  \includegraphics[scale=0.42]{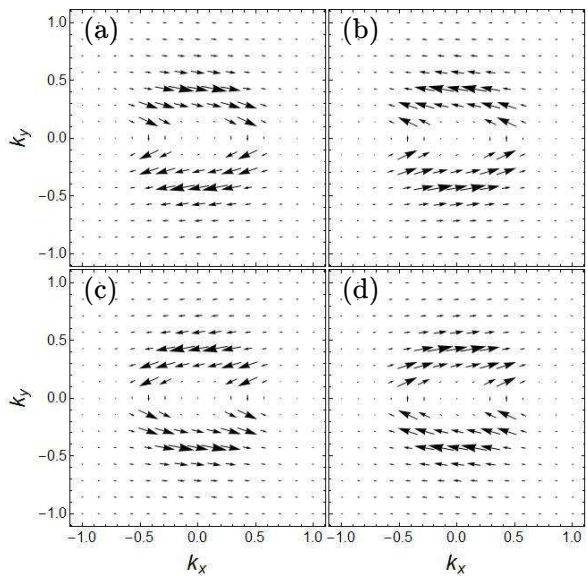}
  \caption{Pseudospin texture of lower [(a) and (c)] and upper [(b) and (d)] band with opposite polarizations of the incident beam with $\eta=+1$ [(a) and (b)] and $\eta=-1$ [(c) and (d)]. The texture reverses as the handedness of the beam is reversed.}  \label{pseudospin}
\end{center}
\end{figure}

Next let us now consider illuminating the line node semimetal by an off-resonant laser beam of frequency $\omega$ polarized in the $yz$ plane. The incident beam will generate a vector potential $\mathcal{A}(t)=\mathcal{A}_{y}\eta\sin \omega t\hat{y}+\mathcal{A}_{z}\sin(\omega t+\phi)\hat{z}$, where $\phi=\pm \pi/2$ for elliptically polarized light, $\phi=0$ or $\pi$ for linearly polarized case. Here $\eta=\pm 1$ represents the left or right handedness of the incident light beam. The full time-dependent Hamiltonian, incorporating the effects of incident light on the spectrum of line node semimetal, can be obtained by the minimal coupling prescription, $k\rightarrow k+e\mathcal{A}(t)$. As an approximation we consider absorption or emission processes involving single photons only. In this situation, the Hamiltonian can be simplified into an effective time-independent one as

\begin{equation}\label{floquet}
H_{\mathrm{eff}}=H+\frac{[H_{-1},H_{+1}]}{\omega}+\mathcal{O}\left(1/\omega^{2}\right),
\end{equation}

\noindent where $H_{\pm1}=\frac{\omega}{2\pi}\int_{0}^{2\pi/\omega}H(t)e^{\pm i\omega t}$. In the large frequency regime the higher order terms are small and the above approximation is reasonable.
We use the continuum Hamiltonian, $H$, for the line semimetal and evaluate the light induced term to be 

\begin{equation}
[H_{-1},H_{+1}] = -2e^{2}\eta\mathcal{A}_{y}\mathcal{A}_{z}b_{xy}vk_{y}\sin\phi \times \begin{pmatrix}0 & 1\\ 1 & 0 \end{pmatrix}.
\end{equation}

This gives the effective photon-dressed Hamiltonian as

%\begin{widetext}

%\begin{equation}
 \begin{align}
 H_{\mathrm{eff}}= & [\epsilon_{0}+a_{xy}(k_{x}^{2}+k_{y}^{2})+a_{z}k_{z}^{2}]I + vk_{z}\sigma_{y} \nonumber \\
 &+ [\Delta\epsilon+b_{xy}(k_{x}^{2}+k_{y}^{2})+b_{z}k_{z}^{2}]\sigma_{z} \nonumber \\
 &- \frac{2e^{2}\eta\mathcal{A}_{y}\mathcal{A}_{z}b_{xy}vk_{y}}{\omega}\sin\phi\sigma_{x}.
\end{align}
%\end{equation}

%\end{widetext}

\begin{figure*}[t]
\begin{center}
  \includegraphics[scale=0.50]{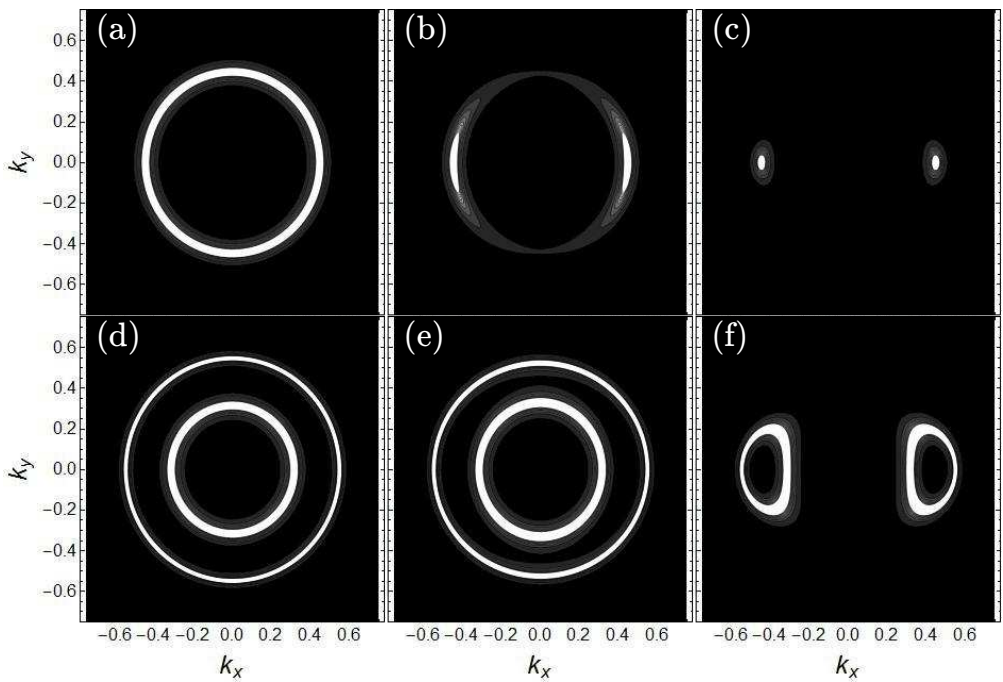}
  \caption{Densities of states in the $k_{x}-k_{y}$ plane at the energy of the nodal line ($E=0$) for (a) $\mathcal{A}_{y}=\mathcal{A}_{z}=0$, (b) $\mathcal{A}_{y}=\mathcal{A}_{z}=0.25$, and (c) $\mathcal{A}_{y}=\mathcal{A}_{z}=0.5$. The same quantity plotted at an energy above the nodal line ($E=0.1$) for (d) $\mathcal{A}_{y}=\mathcal{A}_{z}=0$, (e) $\mathcal{A}_{y}=\mathcal{A}_{z}=0.25$, and (f) $\mathcal{A}_{y}=\mathcal{A}_{z}=0.5$. A broadening of $\delta=0.001$ has been used for these plots.}  \label{dos}
\end{center}
\end{figure*}

Notice that in the effective Hamiltonian, a term linear in $k_{y}$ now appears proportional to $\sigma_{x}$. The energy eigenvalues for $H_{\mathrm{eff}}$ for the case of circularly polarized light are shown in Fig.~\ref{bands}(c). The line degeneracy between the two bands is reduced to a point degeneracy, and a gap is opened at all points except at $k_{y}=0$. The bands are now degenerate only at two points located symmetrically about the center of the circular line node. Thus, the line node semimetal has been converted to a point node semimetal by choosing a suitable polarization and strength of the incident light beam. At the same time the band inversion that we originally started with is still present, unchanged by the incident light. The Hamiltonian which we have chosen obeys $C_{4}$ symmetry, however, this is not crucial for our proposal of light-induced tuning of the line node and similar conclusions can be reached using more general models. 

We note that the light induced term has a $\sin\phi$ dependence, which means that the line node would be perturbed for circularly or elliptically polarized light, while remaining unchanged for linearly polarized case. Although here we have considered the spinless line node semimetal for simplicity, the conclusions also hold for spinful case with strong spin orbit coupling, as can be seen by adding a term $\lambda_{SO}\sigma_{z}\otimes s_{z}$ (here $s_{z}$ represents real spin) to the Hamiltonian. In this case the light induced term is of the form $-2e^{2}\eta\mathcal{A}_{y}\mathcal{A}_{z}b_{xy}vk_{y}\sin\phi\sigma_{x}\otimes I$ and the line node is converted to four point nodes. 

\begin{figure}[b]
\begin{center}
  \includegraphics[scale=0.50]{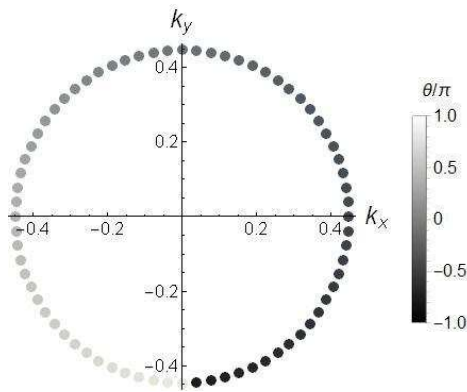}
  \caption{Position of the point node as a function of varying angle of the incident laser beam. The position can be continuously tuned as the beam is rotated.}  \label{nodeposition}
\end{center}
\end{figure}

The effect of incident light on the eigenstates of the system is revealed by calculating the pseudospin, $s=\langle\psi|\sigma|\psi\rangle$. As shown in Fig.~\ref{pseudospin}, the pseudospin vectors originate from and end at the point nodes. The two bands have opposite texture in the $k_{x}-k_{y}$ plane [Fig.~\ref{pseudospin}(a) and (c)]. Interestingly, this can be reversed by reversing the handedness of the laser beam. The reversal of pseudospin by changing $\eta$ to $-1$, is shown in Fig.~\ref{pseudospin}.

Next let us look at changes in the density of states of the semimetals. In Fig.~\ref{dos} we show the $k_{x}-k_{y}$ resolved density of states with increasing light intensity for $k_{z}=0$. When the Fermi energy is located at the energy of the line node ($E=0$), for $\mathcal{A}_{y}=\mathcal{A}_{z}=0$, the projected Fermi surface is a circle. With increasing light amplitude the spectral weight begins to reduce away from $k_{y}=0$, and we find a pair of arc-like features [Fig.~\ref{dos}(b)]. Further increase in laser intensity leaves two points along $k_{y}=0$. With increasing incident light intensity the projected Fermi surface topology changes from a circle to two isolated points, resulting in a Lifshitz transition. We also plot the density of states when the Fermi energy is located slightly above the line node ($E=0.1$) in the lower panels of Fig.~\ref{dos}. In this case there is an additional circular feature around $k_{x}=k_{y}=0$ arising from the second band. Increasing light intensity again leads to a change in the nature of the projected Fermi surface, with the two circular features starting to merge and eventually splitting off to form two disconnected pockets. 

Our results for light-induced conversion of a line node to a point node semimetal can be understood in the framework of general arguments for the stability of line nodes put forth by Burkov, Hook and Balents~\cite{burkov2011topological}. For a general two-band Hamiltonian $H=\sum_{i=x,y,z}h_{i}(k)\sigma_{i}$, nodal lines will be obtained if one of $h_{i}(k)$, is zero for all values of $k$. Then, one needs to tune the remaining two parameters to create the line degeneracy. Using the other two $h_{i}$'s one can construct a complex order parameter. The integral over a closed curve of the phase of the order parameter gives a winding number which is quantized and reveals the presence or absence of the line node. If one of the $h_{i}$ turns out to be finite, then the curve with a finite winding number does not necessarily enclose a node. In our case without illumination with light, it is indeed the situation that there are no terms in the Hamiltonian proportional to $\sigma_{x}$ and we obtain a line node. On the other hand, the photon-dressed Hamiltonian has a term depending on the vector potential strength which is proportional to $\sigma_{x}$ and is not zero for all $k$. As a consequence the line node is no longer stable. The additional term vanishes for $k_{y}=0$, and leaves the degeneracy at two points along this line. As a result the line node semimetal is converted to a point node semimetal. An alternative is to look at the light induced transition from line node to point nodes from a symmetry point of view. The line node is protected by a combination of inversion and time reversal symmetries. Application of light lowers these symmetries, resulting in the loss of topological stability of the line node.

Till now we have fixed the light to be in the $yz$ plane. If we had started with the laser beam polarized in the $xz$ plane the light-induced term would have been of the form $-2e^{2}\eta\mathcal{A}_{x}\mathcal{A}_{z}b_{xy}vk_{x}\sin\phi\sigma_{x}$. Such a term would also gap out the line node except at two points along $k_{x}=0$. This means that we would have obtained a point node semimetal with the nodes located along $k_{x}=0$, i.e., in a direction orthogonal to the case when the laser beam is polarized in $yz$ plane. This observation suggests a direct way to obtain point nodes at different positions in the $k$-space. One needs to simply rotate the laser beam in the plane perpendicular to the plane originally containing the line node. Let us consider the general case when the circularly polarized light beam is incident making an angle $\theta$ with the $k_{x}$ direction. In this situation there are two point nodes created at $(k_{x1}=-\sqrt{\frac{-\Delta \epsilon}{b_{xy}}}\sin\theta,k_{y1}=\sqrt{\frac{-\Delta \epsilon}{b_{xy}}}\cos\theta, k_{z1}=0)$ and $(k_{x2}=\sqrt{\frac{-\Delta \epsilon}{b_{xy}}}\sin\theta,k_{y2}=-\sqrt{\frac{-\Delta \epsilon}{b_{xy}}}\cos\theta, k_{z2}=0)$. The position of one of the point nodes is shown with varying $\theta$ in Fig.~\ref{nodeposition}. The partner point node (not shown) is always present at the diametrically opposite point on the original line node. This way, by rotating the laser beam, one can engineer point nodes at the desired location in the momentum space.

\begin{figure}[t]
\begin{center}
  \includegraphics[scale=0.42]{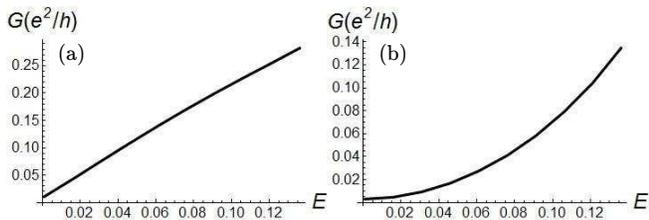}
  \caption{Sharvin conductance as a function of Fermi energy for (a) $\mathcal{A}_{y}=\mathcal{A}_{z}=0$ and (b) $\mathcal{A}_{y}=\mathcal{A}_{z}=0.5$. The scaling dependence on energy changes on applying light.}  \label{conductance}
\end{center}
\end{figure}

There have already been a number of studies where the properties of line node semimetals, as manifested in different measurable quantities, have been contrasted with those of point node semimetals. These propoerties include quantum oscillations~\cite{phillips2014tunable}, Landau level quantization in a magnetic field~\cite{rhim2015landau}, magnetic susceptibility~\cite{koshino2016magnetic}, charge polarization~\cite{ramamurthy2015quasi} and Friedel oscillations~\cite{rhim2016anisotropic}. Our proposal of converting line nodes to point nodes, could be effectively used in conjunction with any of these. Here we point out another discernible signature in the Sharvin conductance. We show the dependence of Sharvin conductance $G$, which is proportional to the number of ballistic channels of the system, on the energy in Fig.~\ref{conductance}. Without light, the conductance varies linearly with $E$, close to the degeneracy point. On applying circularly polarized light, this changes drastically and the variation of $G$ with energy is now quadratic. This is a consequence of underlying change in the density of states for the line node semimetal compared to a point node one. Another observable effect of light induced transition from line node to point nodes can be seen in the Hall conductivity, $\sigma_{ij}$. Due to symmetries Hall conductivity vanishes in the line node case, while becoming finite as well as tunable when light is applied. At low temperatures, for light applied in the $yz$ plane, we find $\sigma_{yz}\varpropto\sqrt{\frac{-\Delta\epsilon}{b_{xy}}}\mathrm{sgn}(\sin\phi)$. In addition to the Sharvin conductance, such a change in Hall conductivity could be measurable in transport experiments.

To get an estimate for the intensities needed to observe our proposal, we set the material parameters, group velocity, $v=5\times 10^{5}$ m/s and coefficient of quadratic term in $k_{y}$, $b_{xy}=1$ eV/\AA{}$^{2}$. Experimentally applied frequencies range in thousands of terahertz, and we choose $\omega=1000$ THz. Then, to open a gap of 20 meV one would need an intensity, $I\sim 10^{12}$ W/m$^{2}$, which appears to be within experimental reach~\cite{wang2013observation,sie2015valley,kim2014ultrafast}.

In summary, we have shown that application of light of appropriate polarization allows turning a line node semimetal into a point node semimetal. These point nodes can be readily controlled by adjusting the applied light beam. Such a light induced change in Fermi surface topology entails changes in a number of measurable quantities, including quantum oscillations, Landau level quantization, magnetic susceptibilty and charge polarization. These could be used in future as concrete tests of our proposal.

\textit{Note added--} After completion of this work, preprints by Yan and Wang~\cite{yan2016tunable} and by Chan, Oh, Han, and Lee~\cite{chan2016type} appeared which also discuss the emergence of Weyl semimetals by periodically driving line node semimetals.

\textit{Acknowledgments--} I would like to thank Victor Chua, Srinidhi Ramamurthy and Smitha Vishveshwara for useful discussions.

\end{document}